\documentclass[preprint,12pt]{elsarticle}

\usepackage{amssymb}
\usepackage{amsmath}
 \usepackage{amsthm}
\usepackage{url} 
\journal{Ocean Engineering}

\begin{document}

\begin{frontmatter}



\title{Flying shape and aerodynamics of a full-scale flexible Olympic windsurf sail}


\author[1]{Jishen Zhang}%

\author[1]{Gauthier Bertrand}

\author[2]{Marc Rabaud}

\author[3]{Benoît Augier}

\author[1]{Marc Fermigier\corref{cor1}}
\ead{marc.fermigier@espci.fr}

\cortext[cor1]{Corresponding author}

\affiliation[1]{organization={Physique et Mécanique des Milieux Hétérogènes, ESPCI, Sorbonne Université, UMR CNRS 7635},
addressline={7bis quai Saint Bernard},
postcode={75231},
city={Paris},
country={France}}

\affiliation[2]{organization={Université Paris-Saclay, CNRS, FAST},
addressline={Bât. 530, rue André Rivière},
postcode={91405},
city={Orsay},
country={France}}

\affiliation[3]{organization={IFREMER, RDT Research and Technological Development Unit},
addressline={1625 route de Sainte-Anne},
postcode={29280},
city={Plouzané},
country={France}}

%% Abstract
\begin{abstract}
The 3D flying shape of a real-scale 8 $\mathrm{m^2}$ iQFOiL class windsurf sail is measured in steady state sailing configurations. The sea wind flow conditions are simulated in a large-scale wind tunnel and stereo camera imaging technique ensure the flying shape reconstruction. Sail form reconstruction allows to measure the twist of the sail profiles from bottom to top. Simultaneous aerodynamic forces and moments applied to the sail are measured via an embedded force balance.  With the measured forces and moments the lift, drag and roll coefficients are determined for various wind intensity. A systematic reduction of these coefficients is observed as compared to previous studies on reduced-scale rigid sail model. We suggest that the sail deformation in the wind is crucial to explain these reductions.
\end{abstract}



%% Keywords
\begin{keyword}
Windsurfing \sep Sail aerodynamics \sep Wind tunnel test \sep Full scale experiment \sep Fluid-structure interaction \sep Photogrammetry



\end{keyword}

\end{frontmatter}



\section{Introduction}
In competitive sailing, the performance of sails is significantly influenced by their structural deformations. Except for some racing classes using rigid wings ~\cite{Fiumara2016WingSail,Giovannetti2024WingSail}, most sails are flexible structures that continuously interact with the surrounding wind flow, creating complex fluid-structure interaction (FSI) problems. This complexity is particularly pronounced in highly cambered downwind sails and deformable rigs, such as those found in windsurfing ~\cite{fagg1997,alexander1998}. Windsurf sails, unlike traditional cruising or racing yacht sails, are designed without the support of stays and shrouds~\cite{hart2014windsurfing}. They are highly flexible and operate as part of a unified rig, with the integrated mast and boom allowing direct control by the rider. Although windsurf sails are nearly inextensible within typical wind speed ranges, the unsupported mast leads to significant deformation of the entire rig under wind loading. This unique characteristic makes windsurfing an excellent example of complex fluid-structure interactions, combining technical challenges for competitions. As such, to optimize a sail's aerodynamic efficiency and overall performance under real sailing conditions, it is crucial to obtain precise measurements of its flying shape~\cite{souppez2024}.

Various experimental techniques have been developed to measure the full-scale flying shape of sails under sea sailing conditions. These methods include Time of Flight (ToF) radar scanning~\cite{Fossati2015}, monocular imaging~\cite{maciel2021monocular}, and multiple synchronized cameras combined with both discrete~\cite{clauss2005cfd,deparday2016full} and continuous~\cite{mausolf2011photogrammetry} pattern recognition algorithms. These techniques offer satisfactory accuracy and enable time resolved 3D reconstructions of the flying shape. However, measuring sail shape in steady-state conditions at sea presents challenges due to the inherently unsteady nature of the wind, both in terms of amplitude and direction~\cite{ViolaFlay2011,Deparday2014,Fossati2015}. To address these challenges, wind tunnel investigations using reduced-scale sail models under controlled wind conditions are commonly employed as alternatives~\cite{Fossati2006,fossati2009aero,graf2009photogrammetric,sacher2017efficient,aubin2018performance}. However, such investigations may lead to aerodynamic-structural discrepancies, including mismatched Reynolds numbers as well as variations in the ratios of fabric weight to wind pressure and membrane stress to wind pressure~\cite{deparday2016full}. To our knowledge, full-scale wind tunnel tests of windsurf sails have not yet been reported.

In the context of windsurfing, limited attention has been given to the joint aerodynamic-structural measurements of real-scale windsurf sails. Alexander and Furniss~\cite{alexander1998} conducted measurements of the aerodynamic coefficients of 1/8-scale rigid models of a Gaastra sail, with the model shapes derived from full-scale sails. Their study specifically quantified the reduction in lift coefficient as the sail was flattened and twisted. Similarly, Mok et al.~\cite{mok2023performance} examined the performance of a 1/4-scale rigid model of an iQFOiL sail~\cite{IQFoil2023} in a wind tunnel. The model was based on full-scale sail shapes rigged indoors without wind, and the study investigated the effects of rig-back angle and Reynolds number on aerodynamic coefficients by varying the angle of attack between 0$^\circ$ and 70$^\circ$. In summary, both studies concentrated on aerodynamic efficiency using reduced-scale rigid models. This approach is not fully representative for flexible rigs for windsurf sails, whose deformations under wind loading can significantly alter aerodynamic performance.

In the following we present wind tunnel measurements of the aerodynamic coefficients of a full-scale Olympic class iQFOiL sail, with simultaneous photogrammetric determination of its 3D flying shape. The paper is structured as follows: we begin by detailing the geometric and mechanical characteristics of the sail, including the rigging configurations used. Next, we describe the experimental setup and the wind tunnel testing conditions. We then introduce the two-camera imaging technique employed for flying shape determination, followed by a validation check of the method. Using the reconstructed 3D sail shape data, we quantitatively analyze the sail twist and mast deformation resulting from aerodynamic loading. Finally, along with the simultaneous force measurements, we provide a comprehensive analysis of the sail's deformability on its aerodynamic performance.

\section{Experimental setup}

\subsection{Sail characteristics}

\begin{figure}[ht]
	\begin{center}
\includegraphics[width=0.5\columnwidth]{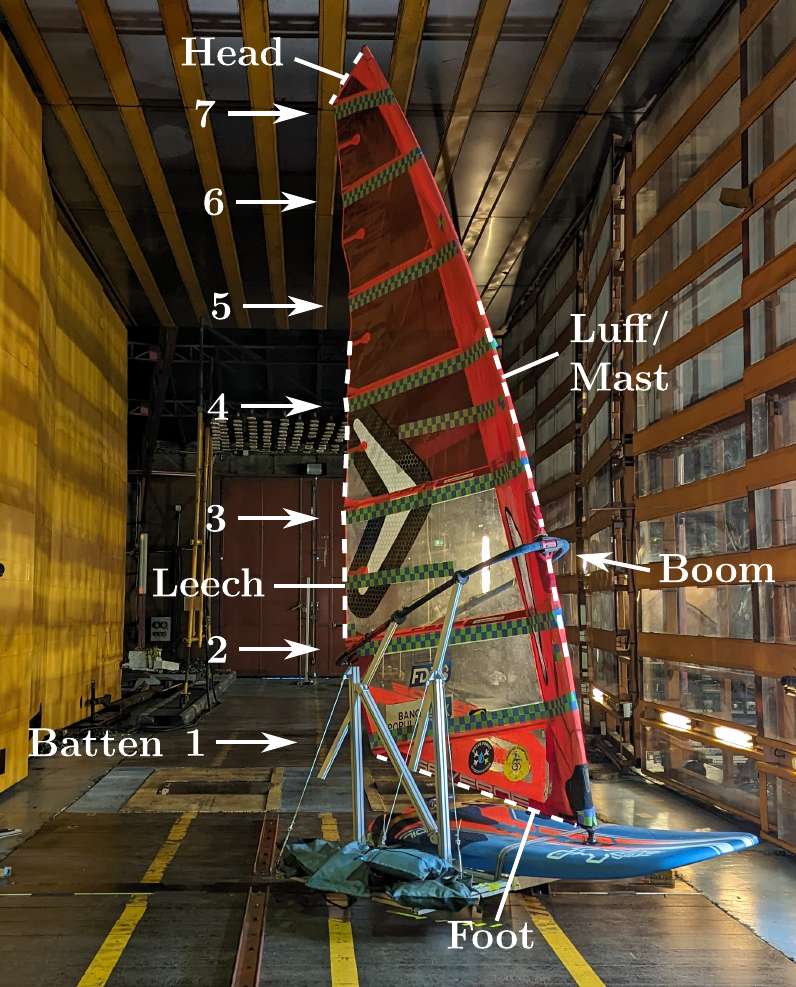}
\caption{Severne HGO 8 m$^2$ sail rigged on a 95 v3 Carbon Reflex board and installed in the S6 wind tunnel at IAT Saint-Cyr l'Ecole. Green/blue checkerboards glued on the sail are used to determine the 3D shape by photogrammetry.}
\label{fig:newPhotoVoile}
\end{center}
\end{figure}

We use a Severne HGO iQFOil 8m$^2$ sail rigged on an Apex 490 carbon fiber mast which is the approved rigging for women in the 2024 Olympic windsurf class. The mast base is flexibly linked to a Starboard iQFOiL 95 Carbon Reflex board (Fig.~\ref{fig:newPhotoVoile}). The mast (4.9 m long) is slid into a sleeve at the luff (leading edge) of the sail and bent by tensioning the luff with a system of pulleys. The maximum chord length $C$ is 2 m.

Seven full battens and 5 battens of reduced-length between the 6 upper ones are distributed along the leech of the sail (trailing edge). The full battens are in compression and bend the sail in an asymmetrical cambered shape, even in the absence of wind. Their tension are adjusted by screws at the trailing edge.

A double-sided carbon fiber boom (wishbone), 2.3 m long, attached to the mast, holds the clew point of the sail with the outhaul line. Adjusting the tension on the outhaul modifies the camber of the sail: a large tension on the outhaul flattens the sail and limits the twist of the sail into the wind while a small tension leads to a larger camber and increases the twist. The settings of the sail were checked by a member of the French national Olympic team. In this study, we will use two different outhaul tensions referred to as low camber and high camber.

\subsection{Wind tunnel}

\begin{figure*}[ht]
	\begin{center}
\includegraphics[width=0.9\textwidth]{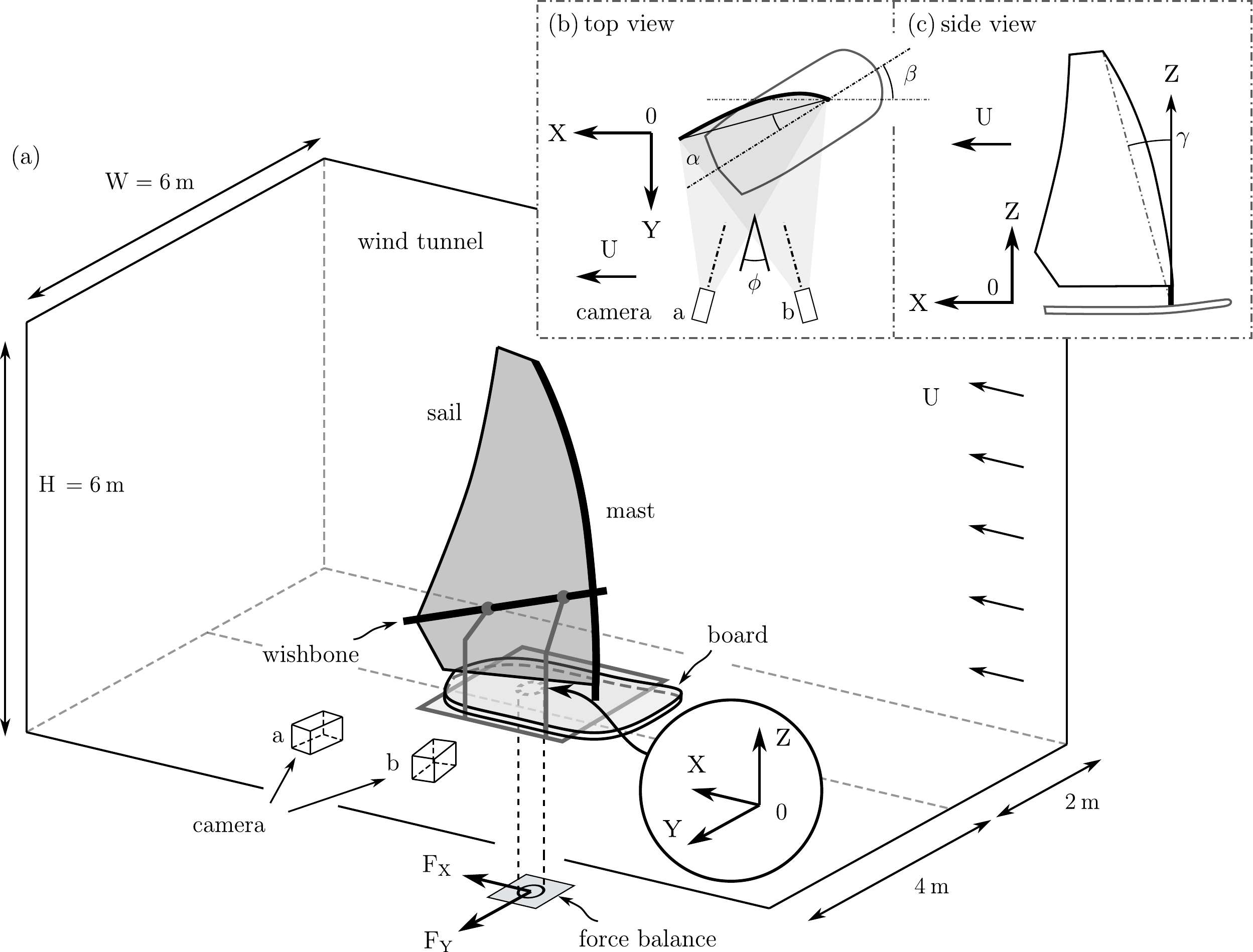}
\caption{Sketch of the full-scale windsurf in the measuring section of the wind tunnel: (a) perspective view, (b) top and (c) side view. The two cameras are fixed on the starboard wall.}
\label{fig:schema_setup}
\end{center}
\end{figure*}

The measurements are carried out in the large low speed S6 wind tunnel at Institut Aérotechnique Saint-Cyr-l'École\footnote{\url{https://iat-en.cnam.fr/}} with a test section measuring $W=6$ m in width and $H=6$ m in height. The tunnel is of open circuit blowdown type with an assembly of 36 fans located 3 m upstream of the test section. The wind in the test section is horizontal and mainly constant along the vertical axis. Turbulent intensity is between 4 and 7 $\%$ which corresponds to a highly turbulent air flow, but similar to the one observed in sailing conditions. At 10 m above sea surface it could be of the order of 10 $\%$ \cite{stull2012introduction}. The free-stream wind speed $U$ is varied in the range [4-8] m/s, so the Reynolds numbers of the flow around the sail, $Re=U C/\nu$ based on the maximum chord $C$, is in the range $0.52\times 10^6$ to $1.04\times 10^6$. Fig.~\ref{fig:schema_setup}a is a sketch of the test bench of the windsurf in the wind tunnel. 

The sail is attached to the board using a universal joint, and its orientation is controlled using two ball joints that keep the wishbone in a fixed position~(Fig.~\ref{fig:newPhotoVoile}). For our measurements, the wishbone's position is set in a way to keep the mast inclined backwards with respect to the board with a medium rig-back angle $\gamma=18\pm1^\circ$ which is kept constant. The board is  fixed to a horizontal carriage mounted atop the force balance, which is positioned beneath the wind tunnel floor. The entire board and sail can be rotated around the vertical axis of the balance to vary the wind incidence angle on the sail.

The $(X,Y)$ plane of our coordinate reference system ($0, X, Y, Z$) is defined as the horizontal plane of the wind tunnel. The $X$ axis is the wind flow direction, $Z$ axis points verticaly upward and $Y$ axis is perpendicular so that the frame is direct. The origin, $0$, is located on the rotating axis of the force balance at the top surface of the board.

The force balance can measure 5 quantities: $F_X$ force component in the direction of wind (i.e. the drag), $F_Y$ horizontal force component perpendicular to the wind direction (i.e. the lift) and $M_X$, $M_Y$ and $M_Z$ moments around the axis $X$, $Y$ and $Z$. The force balance is shifted by one meter from the middle of the tunnel to provide a large enough field of view for the cameras (Fig.~\ref{fig:schema_setup}a). The forces and moments are acquired at a rate of 1 kHz and averaged over a time span of 20 s.

In order to avaluate the aerodynamic forces and moments on the sail, for each measuring point, the forces and moments on the structure and the board, are measured without sail, and subtracted from the raw measurements.

Unlike a rigid 3D profile with zero twist, the sail naturally exhibits a vertical twist, which increases with the wind speed. As a result, a clear definition of how we define the angle of attack (AOA) is required. For that we choose an horizontal plane $Z=67$ cm, that cut the 3D shape of the sail under the wishbone (between battens 1 and 2). The AOA is defined as the angle between the chord of this section and the $X$ axis of the coordinate frame. Thanks to the 3D reconstruction of the 3D sail shape, we are able to determine the \textit{in-situ} AOA with high precision.

To mimic the weight of the rider and to decrease the measured moment $M_X$, counterweights (60 kg of sandbags) are placed upwind on the assembly that holds the board (Fig.~\ref{fig:newPhotoVoile}). Diminishing this rolling moment is essential to get accurate values of the force components.

\subsection{Photogrammetry}

We use a stereo photogrammetry technique to determine the 3D shape of the sail. Two synchronized DSLR cameras (Nikon D800, 7360 $\times$ 4912 pixels and Nikon D750, 6016 $\times$ 4016 pixels, fitted with a Nikon $20~\mathrm{mm}~f/2.8$ lens) are mounted on the wall of the tunnel on the pressure side of the sail. The distance between cameras is 4 m and the angle between their optical axes is $\phi = 53.1^\circ$~(Fig.~\ref{fig:schema_setup}a,b).

Blue-green colored checkerboard strips, aligned with the battens, are glued on the sail (Fig.~\ref{fig:newPhotoVoile}) to provide feature points. Additional small checkerboards are positioned on both leech and luff to locate the sail's contour and the bottom of the mast.
% % \Jishen{We chose to measure the pressure side of the sail surface, 
% as the strong curvature of the suction side at the luff edge would cause 
% significant distortion or even loss of certain feature points, leading to a 
% compromised quality of the reconstructed 3D shape.}}

In order to locate automatically the checkerboard corners in each pair of stereo images, we first transform the color images into greyscale images, taking advantage of the contrast between the sail (essentially red) and the checkerboard strips (blue/green) to reduce background contrast. To identify the checkerboard corners we use an algorithm adapted from Duda \& Reese~\cite{ Duda2018}: we compute the Radon transform \cite{radon20051} of the image and for each image point we calculate the amplitude of angular variation of the transform. Checkerboard corners are characterized by a large value of this amplitude. Thresholding this amplitude map identifies quite effectively the corners. Base on the pixel positions giving the local maximum amplitude, a Gaussian curve fitting is performed around each position allowing the feature points to be determined with a subpixel resolution. A few spurious detected points are eliminated manually. 

Once the corners are identified on each camera view, pairing them and using the calibration of the optical space allow us to reconstruct the 3D coordinates of each detected point in the reference frame of the wind tunnel.

Given the subpixel resolution of the feature point and a scale factor of 0.91 mm/pixel, the location of the points on the sail is determined with an uncertainty on the order of 1~mm in the $X$ and $Z$ directions and a few mm in the $Y$ direction.

%%%%%%%%%%%%%%%%%%%%%%%%%%%%%%
\section{Flying shapes}

\begin{figure*}[ht]
\centering
\includegraphics[width=0.9\textwidth]{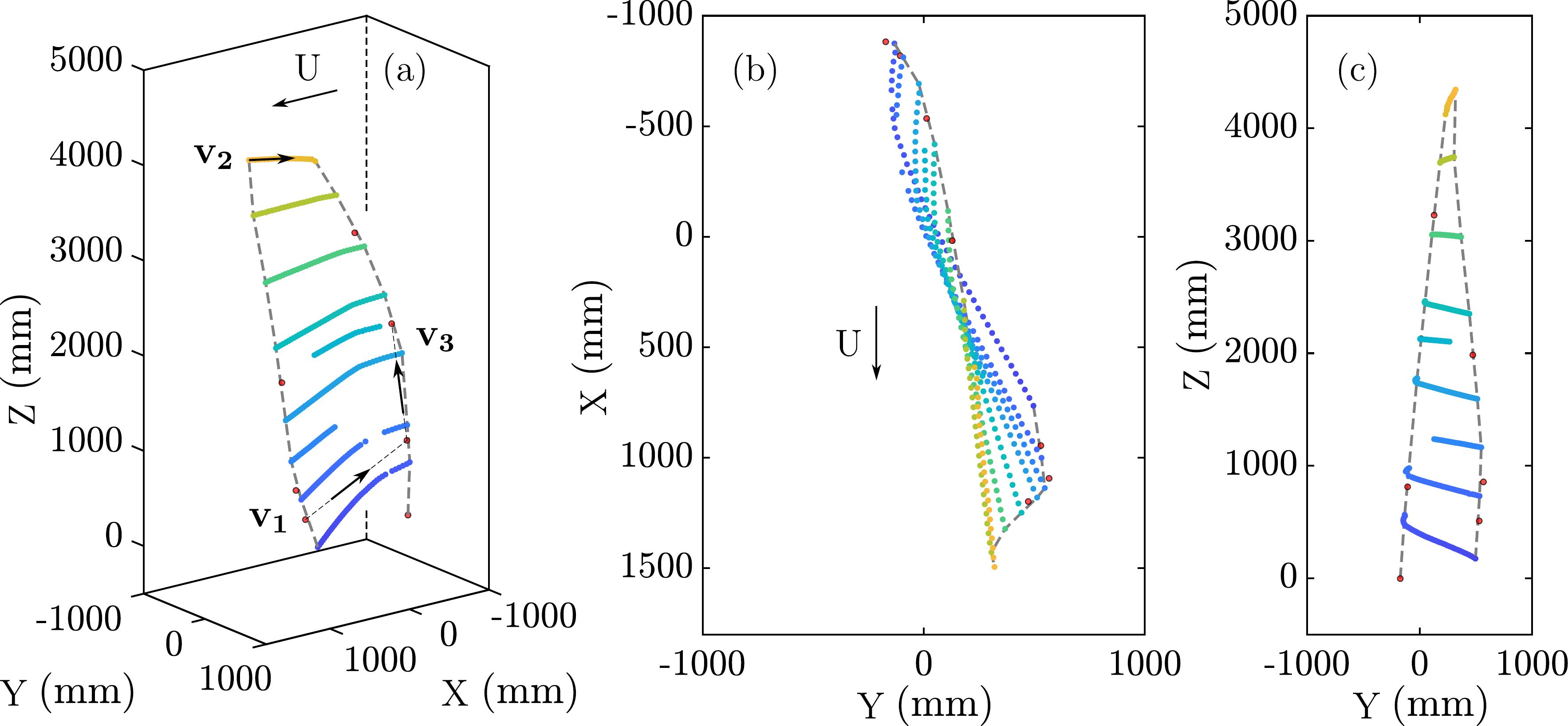}
\caption{Reconstructed shape of the pressure side of the sail in the coordinate system of the wind tunnel, obtained for $U$ = 6 m/s, AOA = 19.8$^\circ$ and high camber case. The color code from blue to orange represents the value of the $Z$ coordinate. The red dots are the additional reference points on the luff, leech and mast base. (a) 3D perspective view. $\mathbf{V_1}$: vector between luff and leech used to determine the angle of attack; $\mathbf{V_2}$: vector along the top batten used to determine the twist; $\mathbf{V_3}$: vector along the lower luff used to determine the mast orientation. (b) top view (along $Z$) and (c) rear view (along $X$).}
\label{fig:24}
\end{figure*}

Using the photogrammetry technique described above, we are able to determine the shape of the sail affected by the wind action, the so-called flying shape. An example of such a flying shape, measured at a wind speed of 6 m/s, angle of attack of 19.8$^\circ$, with a high camber, is shown on Fig.~\ref{fig:24}. In order to characterize the sail shape, we define three vectors: $\mathbf{V_1}$ defined by reference points on the luff and leech between battens 1 and 2. The vector $\mathbf{V_2}$ is along the top batten and vector $\mathbf{V_3}$ between two reference points on the bottom of the mast. The angle between the horizontal component of $\mathbf{V_1}$ and the horizontal component of $\mathbf{V_2}$ defines the global twist of the sail.

\begin{figure}[ht]
\centering
\includegraphics[width=0.25\textwidth]{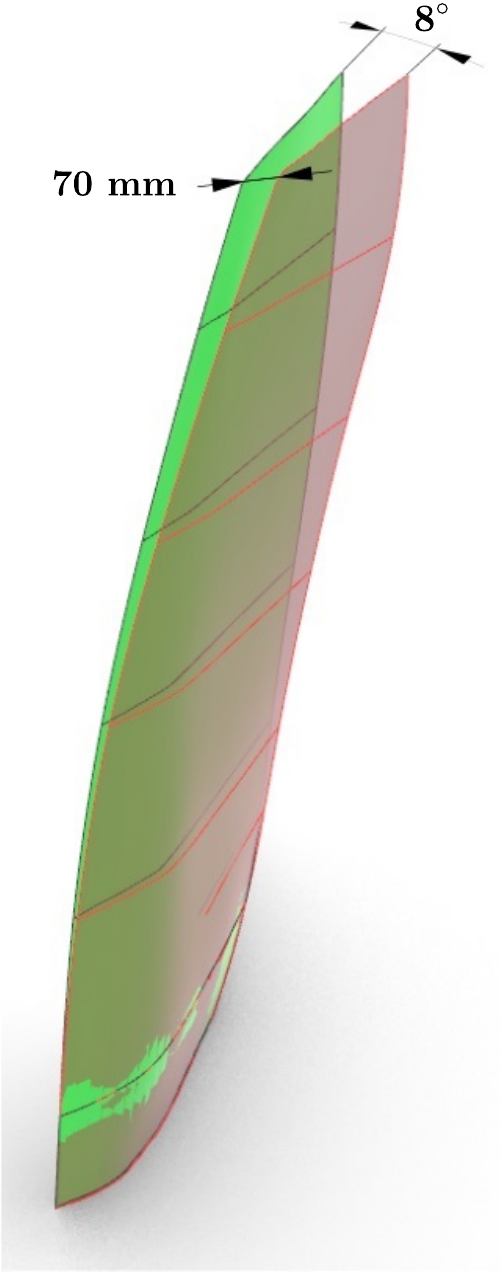}
\caption{Comparison of two 3D reconstructed shapes of the pressure side of the sail obtained for $U$ = 8 m/s in the low camber case: AOA = 1.1$^\circ$ (green) and AOA = 16.1$^\circ$ (red). The bottom of the sail and the bottom part of the mast have been superimposed to highlight the deformation of the top of the sail with the increase of AOA: the twist increases by 8° and the lateral deflection of the masthead by 70 mm.}
\label{fig_Rhino}
\end{figure}

A comparison of two sail shapes  when the angle of attack is increased is presented in Fig.~\ref{fig_Rhino}. Increasing the AOA not only introduces additional twisting in the upper part of the sail but also increases the lateral deflection of the mast.

\begin{figure*}[ht]
\centering
\includegraphics[width=0.9\textwidth]{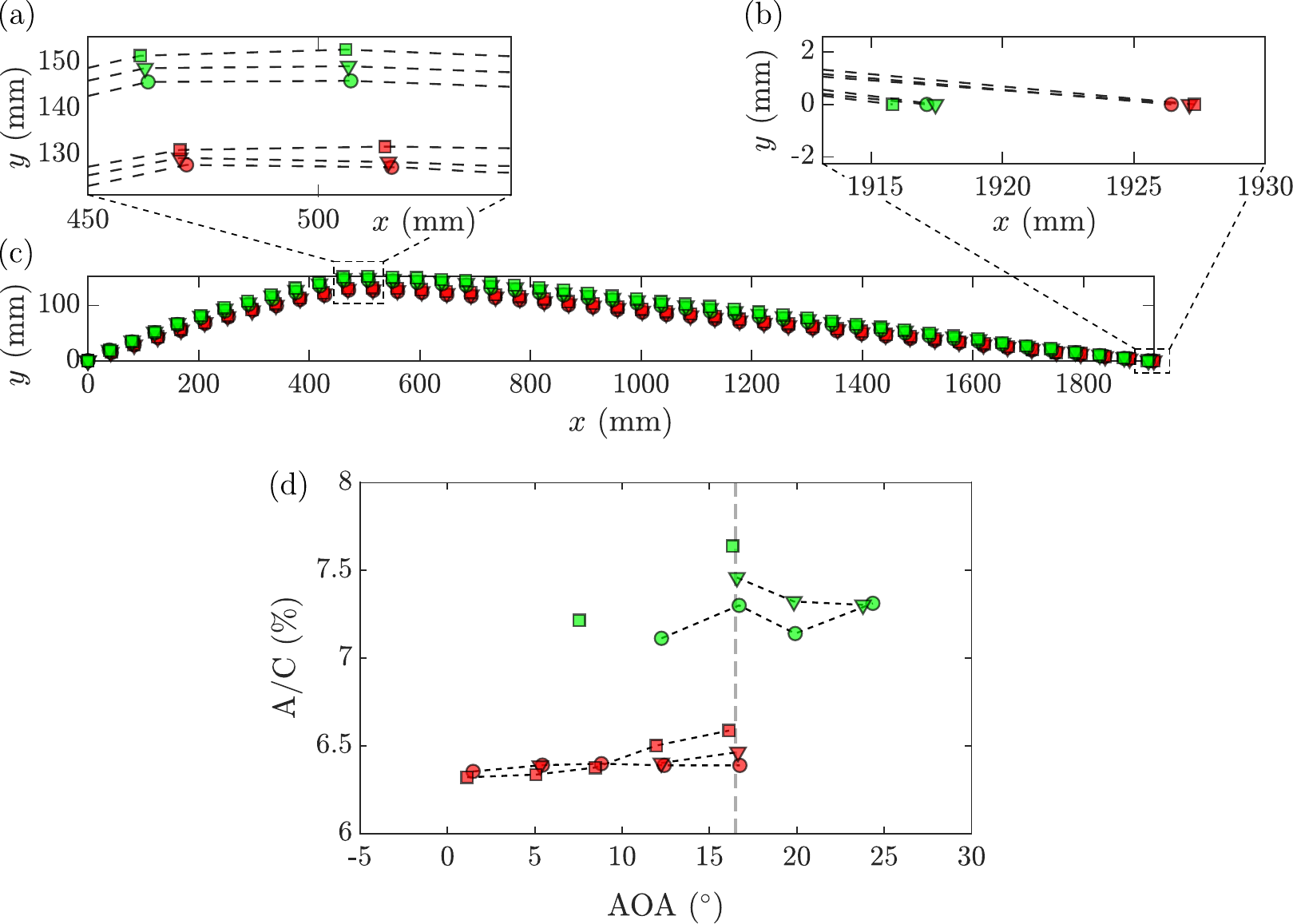}
\caption{Profile of the $3^{\mathrm{rd}}$ batten for AOA = $16.5^\circ$ (red: low camber; green: high camber); $\bigcirc$: $U = 4~\mathrm{m/s}$; $\bigtriangledown$: $U = 6~\mathrm{m/s}$; $\square$: $U = 8~\mathrm{m/s}$. (a) Zoom close to the maximum camber, (b) zoom near the trailing edge, (c) global shape. (d) Maximum camber $A$ normalized by the chord $C$ of the 3rd batten in percent for all the measured AOA. The grey dashed line corresponds to the AOA = $16.5^\circ$ selected for (a), (b) and (c). The profiles are represented in a local coordinate $(0,x,y)$ where (0,0) is the front part of the mast (leading edge of the sail), $x$ along the chord of the profile and $y$ in the transverse direction. By this convention, all profiles are rotated around the leading edge and aligned along the chord $x$ for better quantitative comparison.}
\label{fig:profile}
\end{figure*}

We first focus on the influence of the outhaul tension and of the wind speed on the shape of the sail. To do so, we examine the profile at the level of the 3rd batten. Fig.~\ref{fig:profile} shows this profile for a fixed angle of attack (AOA = 16.5$^\circ$), for the two values of outhaul tension and for three values of wind speed (4, 6 and 8 m/s). The insets on this figure show details of the profiles at the maximum camber and at the leech (trailing edge). The difference in outhaul tension leads to a 10 mm shift in the position of the trailing edge. When the outhaul tension is reduced, the maximum camber of the sails increases by 10 to 15 mm, depending on the wind speed. The influence of the wind speed on the sail shape can be readily seen as the maximum camber of the sail shifts by several millimeters when the wind increases from 4 to 8 m/s.

\begin{figure}[ht]
\centering
\includegraphics[width=0.5\textwidth]{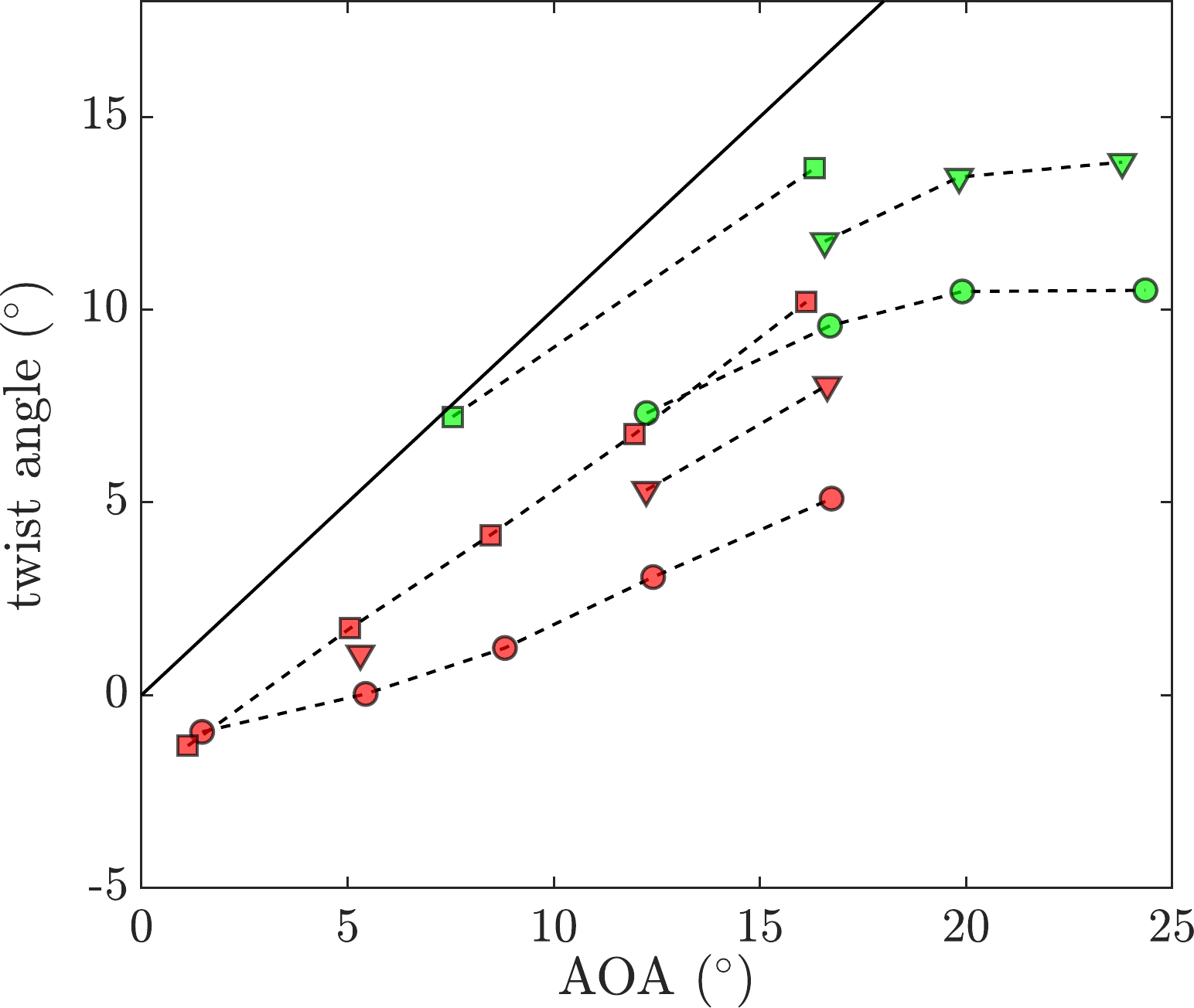}
\caption{Mesured global twist angle of the sail versus the angle of attack (red: low camber; green: high camber; $\bigcirc$: $U = 4~\mathrm{m/s}$; $\bigtriangledown$: $U = 6~\mathrm{m/s}$; $\square$: $U = 8~\mathrm{m/s}$). The black line corresponds to the first diagonal when the twist angle would be equal to the AOA.}
\label{fig:vrillage}
\end{figure}

Another impact of air pressure is the change in sail twist, as shown in Fig.~\ref{fig:vrillage}. When the sail is rigged on the mast and wishbone, it is already slightly twisted, even without wind. However, as the pressure difference between the two sides of the sail increases, the less constrained top of the sail tends to align with the wind direction. A first evidence from Fig.~\ref{fig:vrillage} is that the twist increases with the angle of attack (AOA). A larger AOA results in a greater pressure difference and increased structural stress on the rig. The rig responds by deforming in its easiest mode, the torsional mode, with the secondary mode being the bending of the mast in the plane perpendicular to the wind. For a given AOA, the high camber setting produces a larger twist angle compared to the low camber case, a phenomenon confirmed by practical field experience. In Fig.~\ref{fig:vrillage}, the continuous line corresponds to a top of the sail aligned with the wind. All measured points fall below this line, indicating that the twist effect reduces the angle of attack but keeps it positive, even at the sail top. Consequently, aerodynamic lift is generated across the entire sail section.
Note that two twist angles of the low camber sail are negative for small AOA. This can be attributed to the leech's loose tip, which drops under gravity, here with a small negative incidence of the top of the sail.

\begin{figure}[ht]
\centering
\includegraphics[width=0.5\textwidth]{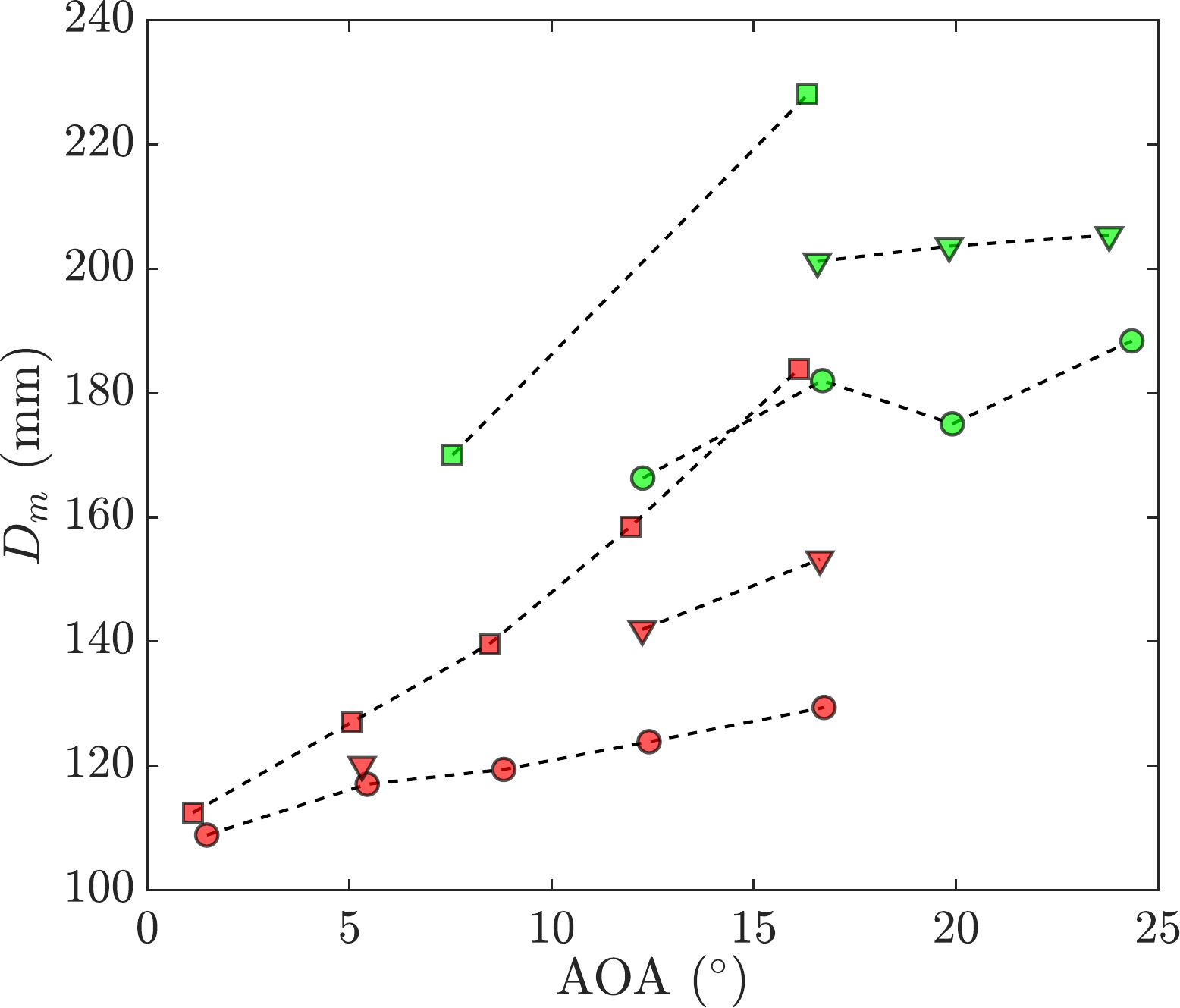}
\caption{Lateral position of the mast head $D_m$ compared to plane defined by the lower part of the sail, as a function of AOA and for different wind speeds and cambers (red: low camber; green: high camber; $\bigcirc$: $U = 4~\mathrm{m/s}$; $\bigtriangledown$: $U = 6~\mathrm{m/s}$; $\square$: $U = 8~\mathrm{m/s}$).}
\label{fig:devers}
\end{figure}

In addition to the twist, the stress due to the wind induces a lateral deflection of the mast towards the leeward side. This is well known by windsurfers, but we can measure this effect as shown on Fig.~\ref{fig:devers}. To eliminate an overall motion of the rig and take into account only this deformation, we determine the mast head lateral position $D_m$ as the distance between the top of the mast and the plane defined by the bottow of the sail (i.e. defined by two vectors $\mathbf{V_1}$  and $\mathbf{V_3}$ (Fig.~\ref{fig:24}a). The evolution of $D_m$ follows the same trend as the twist angle: it increases with AOA and, for a given AOA, it increases with the wind speed. For example, at AOA 16.5$^\circ$ for the low camber case, the mast head moves 6 cm to the leeward side when the wind goes from 4 to 8 m/s. An amplitude on the order of 1~\% of the mast length.

From this analysis, we understand how the whole rig, including sail and mast, deforms. As we report in the following section, this deformability limits the aerodynamic force generated, helping the athletes to sail in strong winds. We expect also a lowering the point of application of the resulting sail force, thus limiting the capsizing torque.

%%%%%%%%%%%%%%%%%%%%%%%%%%%%%%
\section{Aerodynamic coefficients}\label{sec:aero}

\begin{figure*}[ht]
\centering
\includegraphics[width=1\textwidth]{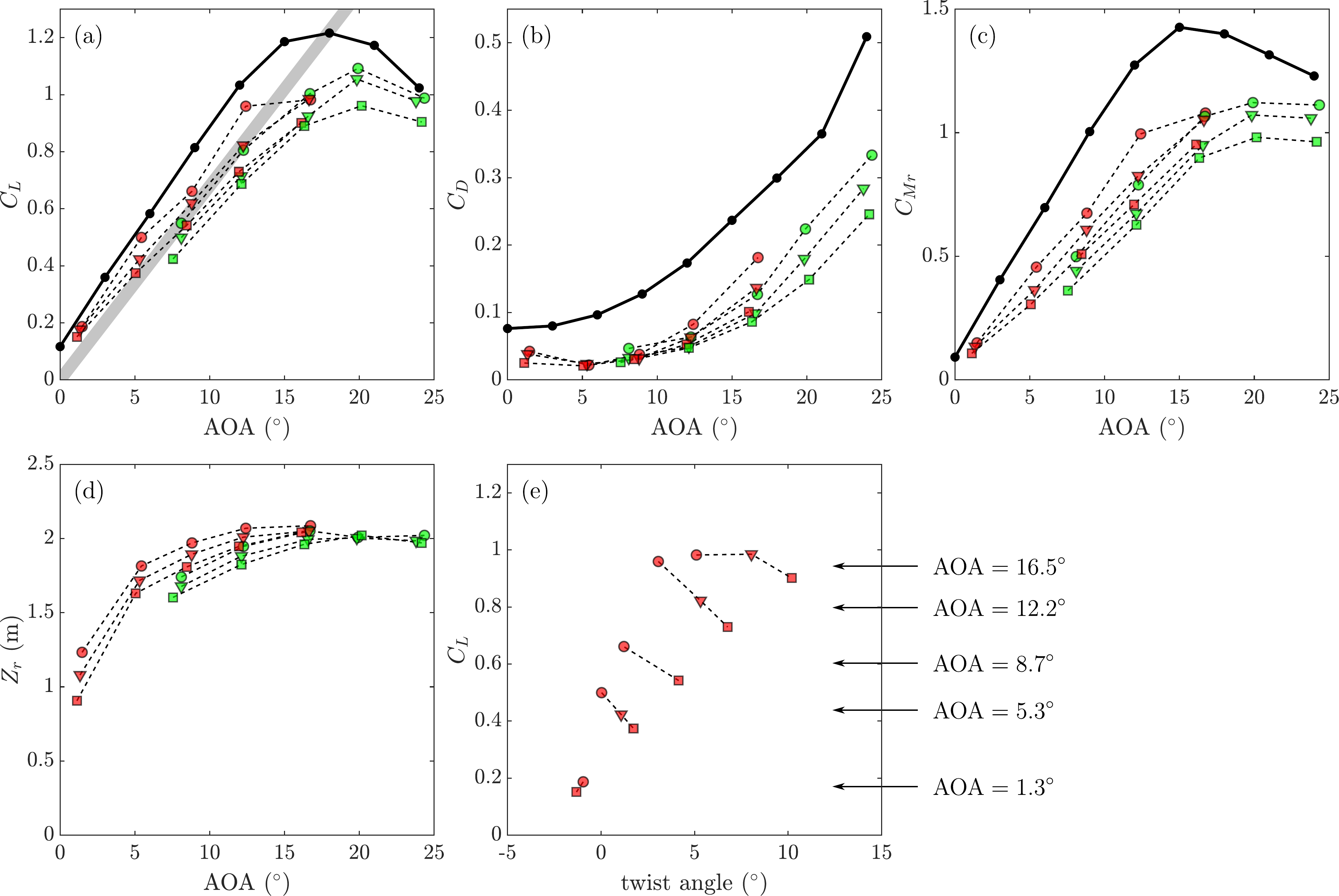}
\caption{Evolution of the aerodynamic coefficient with AOA: (a) lift coefficient $C_L$ (the grey shaded bar corresponds to the classical prediction for a finite span thin symmetrical profile with elliptical loading $C_L=2\pi \lambda/(\lambda+2)\times AOA$ where $\lambda=3.36$ is the present sail aspect ratio), (b) drag coefficient $C_D$, (c) Rolling moment coefficient $C_{Mr}$ versus the AOA and (d) altitude of the centre of pressure $Z_r$ versus AOA and (e) lift coefficient $C_L$ as a function of twist angle. In all the figures, red symbols are for low camber; green for high camber; $\bigcirc$: $U = 4~\mathrm{m/s}$; $\bigtriangledown$: $U = 6~\mathrm{m/s}$; $\square$: $U = 8~\mathrm{m/s}$. Black dots and thick continuous black line are corresponding data from~\cite{mok2023performance}.}
\label{fig:ClCdCm}
\end{figure*}

Together with the sail shapes, the forces and the moments are simultaneous measured for various wind intensities and directions. Only the lift, drag and rolling moment coefficients are analyzed in the present study.
The main effect of the wind intensity is taken into account by dividing these quantities by the dynamic pressure which leads to the usual aerodynamic coefficients :

\begin{equation}
C_L = \frac{F_Y}{\frac{1}{2}\rho U^2 S},~C_D = \frac{F_X}{\frac{1}{2}\rho U^2 S},~C_{Mr} = \frac{M_r}{\frac{1}{2}\rho U^2 S C}
\label{eq:lift_drag}
\end{equation}

where $U$ is the free-stream air flow velocity, $S$ the reference sail surface and $C$ the reference chord of the sail. The rolling moment $M_r$, is defined as the moment of the aerodynamic force $M_r = \sqrt{M_X^2+ M_Y^2}$, were $M_X$ and $M_Y$ are the moments measured in the reference frame of the wind tunnel (Fig.~\ref{fig:schema_setup} b). We also compute the altitude of the center of application of the rolling force $Z_r$ dividing the rolling moment by the aerodynamic force $F_a=\sqrt{F_X^2+ F_Y^2}$:  $Z_r= M_r/F_a$.
Measurements of these quantities, corresponding to both outhaul tensions are shown, for three wind speeds, on Fig.~\ref{fig:ClCdCm}. 

Fig.~\ref{fig:ClCdCm}a shows a classical behavior with an increase of the lift coefficient when AOA increases then a decrease when stall occurs. Low camber case (red symbols) induces a slightly larger lift coefficient compared to a high camber (green symbols). This somehow differs with what might be expected of the influence of camber: a more cambered 2D profile should lead to a higher lift coefficient. However we have seen that the high camber sail is also more twisted than the low camber one (Fig.~\ref{fig:vrillage}) so that the highly twisted shape could be responsible for the decrease of lift.

On the other hand, our data reveal also a wind speed influence on the lift coefficient. For a given AOA, $C_L$ decreases with increasing wind speed for both cases. Generally speaking, in high turbulent flow regimes such as is our case ($Re=5.1\times10^5 - 1.0\times10^6$), $C_L$ is expected to be insensitive to $R_e$. This sensitivity in wind speed suggests that the sail's 3D twist effect must come into play. Contrary to a rigid airfoil with an AOA independent of the span-wise coordinate, the twisted sail profile results in an AOA decreasing from bottom to top. This twist angle being the result of a balance between the wind dynamic pressure and the elastic stresses within the rigging structure, alters also with the wind speed. A higher wind speed leads to a higher twist of the sail, a smaller local AOA and thus a smaller $C_L$. This argument is supported by comparing with experimental data of a 1/4 reduced-scale twist-free rigid sail model, at a similar yet smaller Reynolds number, by Mok et al.~\cite{mok2023performance} (black dots). Our data correspond to smaller values of $C_L$, as a result of the twist-induced local AOA decrease. The difference, however, is smaller at low AOA, possibly due to the smaller forces and, consequently, reduced deformations. The sail twist also induces a stall delay, from around $17^\circ$ for the twist-free rigid sail model, to $20^\circ$ for our case.

The presence of an obstacle in the measurement section of a wind tunnel causes an acceleration of airflow on both sides, known as the blockage effect~\cite{barlow1999low}. This effect can alter the aerodynamic and structural measurements of the sail and is quantified by the blockage ratio, which is the ratio of the sail's projected area to the cross-sectional area of the wind tunnel. If the blockage ratio is less than 10~\%~\cite{spera1994wind,chen2011blockage}, the additional effect can be considered negligible. Given the  projected area of the sail at a large angle of attack (AOA) of $25^\circ$, the blockage ratio is calculated to be 9~\%. Thus, we will neglect this effect in the following. Another effect to consider is the wall effect. A small distance between the wall and the suction side of the sail can induce additional lift and drag~\cite{rosenhead1931,green1947two}. In our study, the sail was shifted 1 m towards the wall to optimize the cameras' field of view, resulting in a 2 m distance between the suction side of the sail and the wall. To estimate this effect, we apply Rosenhead's analysis of potential flow around an inclined 2D flat plate confined between two walls. Based on our data, the additional lift accounts for 33~\% of the total measured lift at an angle of attack (AOA) of $25^\circ$ and should not be neglected. However, these analytical estimations are based on a 2D profile using a potential flow approach, which clearly fails to account for the complexities of a 3D sail, especially with twisted profiles that result in varying AOA along the sail's vertical positions in turbulent flow. In our study, we provide an order of magnitude estimation without applying further corrections to the results obtained from the force and moment measurements.

Data of the drag coefficient are shown in Fig.~\ref{fig:ClCdCm}b. For small wind speeds (AOA $<10^\circ$), all measured data for both rig settings collapse onto a single curve and reach a minimum around $6^\circ$ for the high camber case. Above AOA = $10^\circ$, $C_D$ increases with increasing AOA. 

The data dispersion in wind speed for $C_D$ is more pronounced than for $C_L$. $C_D$ decreases with increasing speed and a maximum reduction of $39~\%$ is found at AOA = $20^\circ$ between two extreme wind speeds. As a matter of fact, a higher wind speed leads to a higher twist and the sail's upper part becomes more aligned to the air flow direction thus reducing the drag. However, when compared with those of the 1/4 twist-free rigid model, the present values $C_D$ are surprising small and even smaller in some cases than the minimum possible value corresponding to the induced drag of a wing with an elliptic lift distribution~\cite{anderson2005introduction}. We can't rule out the possibility of errors in the process of subtracting the drag on the structure and on the board or induced by large transverses values induced by an imperfect diagonalization of measure of the force balance.

The roll moment coefficient $C_{Mr}$ is shown in Fig.~\ref{fig:ClCdCm}d. Similar trends as for $C_L$ are observed: a decrease of the coefficient when the wind speed or the camber increase. 

To highlight the influence of the sail deformation on the roll moment, we also plot on Fig.~\ref{fig:ClCdCm}e the altitude $Z_r$ of the center of effort. We see that data for both cambers follow a similar trend with AOA. The stall-induced lift loss seems to have little impact on $Z_r$, as all data converge to a height of 2 m approximately. Generally speaking, the low camber case leads to a smaller $Z_r$ as compared to the high camber case. The difference reads 20~cm at the optimal AOA around $10^\circ$. At higher AOA, the gap decreases, suggesting that the effect of camber vanishes at large wind angles. The Reynolds number on the other hand has an inverse effect on the center position as expected: a higher $R_e$ leads to a smaller $Z_r$, attributed to an additional twist effect. This dependence becomes less important for large wind angles.

Finally, Fig.~\ref{fig:ClCdCm}e illustrates the variation of lift coefficients as a function of twist angles, for the case of low camber. A general positive correlation between the lift coefficient $C_L$ and twist angle is observed, corresponding to an increase in AOA from 1.3$^\circ$ to 16.5$^\circ$. However, for a constant AOA, increasing wind speed results in a negative correlation between $C_L$ and twist angle: higher wind speeds induce greater twist angles, which in turn lead to lower lift coefficients. At the highest AOA of 16.5$^\circ$, this effect diminishes, likely due to stall occurrence.

% On est capable de mesurer et relier des différences de performances aux diff de géométries image 7 
% On observe 4 effets IFS du vrillage qui diffèrent de l'approche rigide de Mok et al
% Le vrillage augmente avec la vitesse du vent parce que les efforts sont simplement en U² (l'adim ici cache l'effet simple de l'effort)
% Le vrillage augmente avec l'AoA
% le vrillage diminue les efforts aérodynamiques
% le vrillage modifie le point d'application de la résultante Aéro (pour moi, le vrillage fait baisser le centre de poussée !! Contraire à la Fig.~7(f) donc
% La tension dans le outhaul modifie le profil mais également la réponse de la structure ie le vrillage
% là les résultats sont contre intuitif, je m'attends à
% +tension de outhaul - de vrillage --> j'ai demandé confirmation à Pierre Noesmoen
% +tension de outhaul - de creux - de portance}

%%%%%%%%%%%%%%%%%%%%%%%%%%%%%%
\section{Discussion}

Due to the finite width of the wind tunnel, the measurements of the flying shapes were conducted only on the pressure side surface of the sail. The suction side has not the same shape close to mast, as the sail has there a double skin that surrounds the mast. It is only further downwind that the sail becomes a single layer of Mylar. However, the measurement of the pressure side surface alone is an effective method of highlighting the deformations in the sail caused by the flow.

In real sailing condition, the wind is not of constant velocity in the first meters above the see level because of the turbulent boundary layers. As the board moves in a direction different than that of the wind, the apparent wind in the frame of the board is not of constant direction and intensity in $Z$, it is twisted and the AOA increases with altitude for a non twisted profile. This effect can be larger than 5$^\circ$ for a windsurf sailing upstream at 15 knots in 15 knots of true wind.
% \sout{\MarcR{J'ai fait une estimation avec un bas de voile à 1 m et le haut à 6 m, et une rugosit" de surface de surface de 10 cm. Si le vent réel est de 15 N en haut de voile il est de 8.4 N en bas. Si on suppose de plus que la planche avance à 15 N à 45$^\circ$ du vent réel on trouve un angle apparent en haut de mat de 22.5$^\circ$ et en bas de mat de 15.9$^\circ$, soit un vrillage du vent apparent de 6.6$^\circ$.)}}
Apparent wind twist, which depends on boat speed, is quite difficult to simulate in a wind tunnel~\cite{flay1996twisted, richards1997effect}. It should logically increase sail twist at sea compared to wind tunnel measurements.

In this study we coupled measurements of the shape of the sail with measurements of the associated aerodynamic forces. We see through analyses on lift force, drag force and roll moment the determining effects of the sail twist on the sail performance. Both large camber and large wind speed lead to an increase in sail twist, an effect that cannot be analyzed with rigid sail measurement. 
Concerns have been raised regarding the measurement of drag forces, which yielded values of aerodynamic finesse that looks too large when comparing to existing literature but are nevertheless within the order of magnitude for a system with a low aspect ratio, such as a sail.

%%%%%%%%%%%%%%%%%%%%%%%%%%%%%%
\section{Conclusion}

A full scale Olympic windsurf sail has been tested in a wind tunnel. A three-dimensional sail shape reconstruction system has been developed, based on a method of stereophotography detection of points by the Radon transform. The global estimated resolution is on the order of a few millimeters. The effect on sail deformation of various sailing condition and sail parameters was tested, including flow speed, sail angle of attack, and sail tension setting at the clew. These shape measurements were coupled with drag force, lift force, and roll moment measurements to determine aerodynamic performance, taking into account fluid-structure interactions. It was observed that decrease of trailing edge tension and an increase flow velocity resulted in enhanced sail twist and lateral deformation of the mast. In return, elastic deformations of the sail result in modification of aerodynamic performance. This represents a significant performance constraint that windsurfers must consider. Fluid-structure interaction should also be accounted for when testing or simulating such deformable riggings.

%\begin{Acknowledgments}
\section*{Acknowledgments}
We thank Louis Giard, member of the French national sailing team, for rigging and adjusting properly the sail shape and national coach Pierre Noesmoën for fruitful discussions. We thank Clodoald Robert and the technical staff of IAT, as well as Paul Iachkine from the Ecole Nationale de Voile et des Sports Nautiques for their help during the measurements.

This work was supported by the French National Research Agency with grant ``Sport de Très Haute Performance'' ANR 19-STHP-0002.

\bibliographystyle{elsarticle-num}
\bibliography{Biblio_iqfoil}

\end{document}